\documentclass[11pt, oneside]{article}   	% use "amsart" instead of "article" for AMSLaTeX format
\usepackage{geometry}              		% See geometry.pdf to learn the layout options. There are lots.
\usepackage{color} 
\usepackage{colortbl} 
\usepackage{booktabs}
\usepackage{longtable}
\usepackage{rotating}
\usepackage{soul}
\usepackage[all]{xy}
\usepackage{caption}
\usepackage{graphicx}				% Use pdf, png, jpg, or eps§ with pdflatex; use eps in DVI mode					% TeX will automatically convert eps --> pdf in pdflatex		
\usepackage{amssymb}
\usepackage{amsmath}
\usepackage{hyperref}
\usepackage{enumitem}
\newcommand{\DKL}[2]{\textrm{D}_\textrm{KL} \left( #1  || #2 \right) } % Kullback-Leibler-Divergence
\newcommand{\MI}[2]{\textrm{I}(#1;#2)} % Kullback-Leibler-Divergence

\title{%Localizing in Infinity. %
Attracting Sets in Perceptual Networks} 
\author{Robert Prentner\\robert.prentner@gmail.com}
\date{}							% Activate to display a given date or no date
\begin{document}
\maketitle 
\vspace{-0.66cm}
\begin{footnotesize}
%\noindent Address blabla.\\
%$4$ To whom correspondence should be directed. ORCID: 0000-0003-1890-0827
\end{footnotesize}
%\section{}
%\subsection{}
\setstcolor{red}
\begin{center}
\subsection*{Summary}
\end{center}
%
%...\\
%\noindent \textbf{Keywords}: perceptual networks, mutual information on
%\newpage
%
%\section{Introduction}
This document gives a specification for the model used in [1]. It presents a simple way of optimizing mutual information between some input and the attractors of a (noisy) network, using a genetic algorithm. The nodes of this network are modeled as simplified versions of the structures described in the ``interface theory of perception'' [2]. Accordingly, the system is referred to as ``perceptual network''. 

\noindent The present paper is an edited version of technical parts of [1] and serves as accompanying text for the Python implementation \textsf{PerceptualNetworks}, freely available under [3]. 
\\

\begin{itemize}[leftmargin=0.5cm]
\item[1.] Prentner, R., and Fields, C.. Using AI methods to Evaluate a Minimal Model for Perception. \textit{OpenPhilosophy} \textbf{2019}, 2, 503-524.

\item[2.] Hoffman, D. D., Prakash, C., and Singh, M.. The Interface Theory of Perception. \textit{Psychonomic Bulletin and Review} \textbf{2015}, 22, 1480-1506.

%\item Hoffman, D. D., and Prakash, C.. Objects of consciousness. \textit{Frontiers in Psychology} \textbf{2014}, 5:77.

\item[3.] Prentner, R.. \textsf{PerceptualNetworks.} 
\url{https://github.com/RobertPrentner/PerceptualNetworks}. (accessed September 17 2020)
\end{itemize}
\newpage
\section{Introduction}
\subsection{Computational Framework}

The model which we propose in this paper, \textsf{perceptual networks}, combines many features of well-known modeling paradigms but gives them a more psychological interpretation. More concretely, \textsf{perceptual networks} are defined as networks of individual and excitable agents that (i) have experiences corresponding to their states, (ii) are located on a graph which encodes their interaction-topology, and (iii) whose individual actions affect the performance on the network level. Such agents perceive certain messages and send messages, based on their experiences (states), to other agents in their environment. 

Similar to artificial neural networks, \textsf{perceptual networks} can learn from experience by either changing their updating rule or by adjusting their connectivity within the network. These two processes correspond to a ``perceptual'' learning process (i.e. how to represent incoming information) and an ``action-related'' learning process (i.e. where to send information to) respectively. The analysis of information processing capacities embodied by a \textsf{perceptual network} is afforded by its basic graph-structure. \textsf{Perceptual networks} share this feature with random Boolean networks; and the idea that processes on the individual level affect the global functioning of the network is consistent with the main thrust of agent based models and cellular automata. 

We will demonstrate in the following sections how \textsf{perceptual networks} could address two different but related problems. First is the problem of how such a model gives rise to stable structures. 
%Given the more radical reading advanced above, this problem is an ontological one: If there is no real ontology underlying whatever we study, then how is it that an ontologically transient starting point (a network of agents which occasionally get excited) could give rise to a stable structure in the world? 
Second is the problem of how these structures enable ``perception'', if perception is understood broadly in terms of a representation of some input, which results from the application of (internal) rules embodied in the network under conditions of uncertainty (modeled as noise). 

\textsf{Perceptual networks} are a simplification of a framework previously proposed to study consciousness \cite{Hof14}. Given an external measurable space (a ``world''), a ``conscious agent'' in this framework is a six-tuple consisting of two measurable spaces (``perception space'' and ``action space''), three kernels that connect the world and these spaces (``perception'', ``decision'' and ``action'') and a time-counter which counts the number of kernel executions. This definition was refined and a network consisting entirely of ``reduced conscious agents'', a concept introduced to distinguish ``internal'' aspects of a conscious agent from its ``extrinsic'' aspect, was introduced \cite{Fie18}. There it was shown that such a network could be applied to problems in psychology and formally reproduce the architectures of many received models in cognitive science. 

\textsf{Perceptual networks} are simplifications of this more general formalism and feature nodes (``agents'') that integrate information (``messages'') and relay this information to neighboring nodes in the network. 
The mapping of sensory information could be represented by a simple rule $R_j$ which connects the state $m_j$ of a message each agent receives from its neighbors to its future state $x_j$. The state of $m_j$ is determined by the topology of the network $A$:

\begin{equation}
\begin{pmatrix}
x_1\\x_2\\\vdots\\x_n 
\end{pmatrix}
\overset{A}{\longrightarrow} 
\begin{pmatrix}
m_1\\m_2\\\vdots\\m_n
\end{pmatrix}
\overset{\{R_j\} }{\longrightarrow}
\begin{pmatrix}
x_1'\\x_2'\\\vdots\\x_n' 
\end{pmatrix}
\end{equation}

For simplicity, we look at deterministic systems, but the model should eventually be extended to model probabilistic inferences. More precisely, the updating process would then correspond to a homogeneous discrete-time Markov chain, which would feature at least one stationary distribution (under the assumption that it is irreducible and non-periodic), cf. Chapter 3 in \cite{Geb15}.
The construction allows for learning; learning could affect the rules describing information integration, $\{R_j \}$, as well as network topology $A$. 

\subsection{Asymptotic States and Attractors}
\label{sec:example}
The rules define a circular updating scheme from network states to network states. We illustrate the state evolution by a toy network comprising $N=4$ nodes, as shown in Table \ref{tab:one}. The continuous and dotted lines represent two different topologies $A$ and $A'$ (left and middle column), giving rise to two different state evolutions, each starting in the same initial state $(1,0,1,0)$ (right column). Since the network is finite and closed, the state evolution will either land in a stationary state or an attracting set. In the present example an attracting set will be reached in each case.
We have used a simple updating rule which takes the form: 
\begin{equation}
\label{eq2}
m_j = (\textbf{A}\vec{x})_j\ \ \ x_j' = \begin{cases} 1 \textrm{ if } m_j \ge 1 \\ 0 \textrm{ else}\end{cases},
\end{equation}
where $\textbf{A}$ is the adjacency matrix of $A$. Note that in the actual implementation \textsf{PerceptualNetworks} we use a more refined threshold condition than in the present example (cf. Eq \eqref{eq:ten} in \ref{sec:methods}).
\begin{table}[htb]
\centering
\caption{Example of a \textsf{perceptual network} with two different topologies (continuous and dotted lines respectively). Initially, nodes 1 and 3 are excited, the rules (given in text) and the adjacencies of the network (middle column) determine the state evolution (right column).}
\label{tab:one}
\begin{tabular}{ccc}
\\\toprule
Network & Adjacencies & State Evolutions\\
\begin{minipage}{0.33\textwidth}
\includegraphics[scale=0.4]{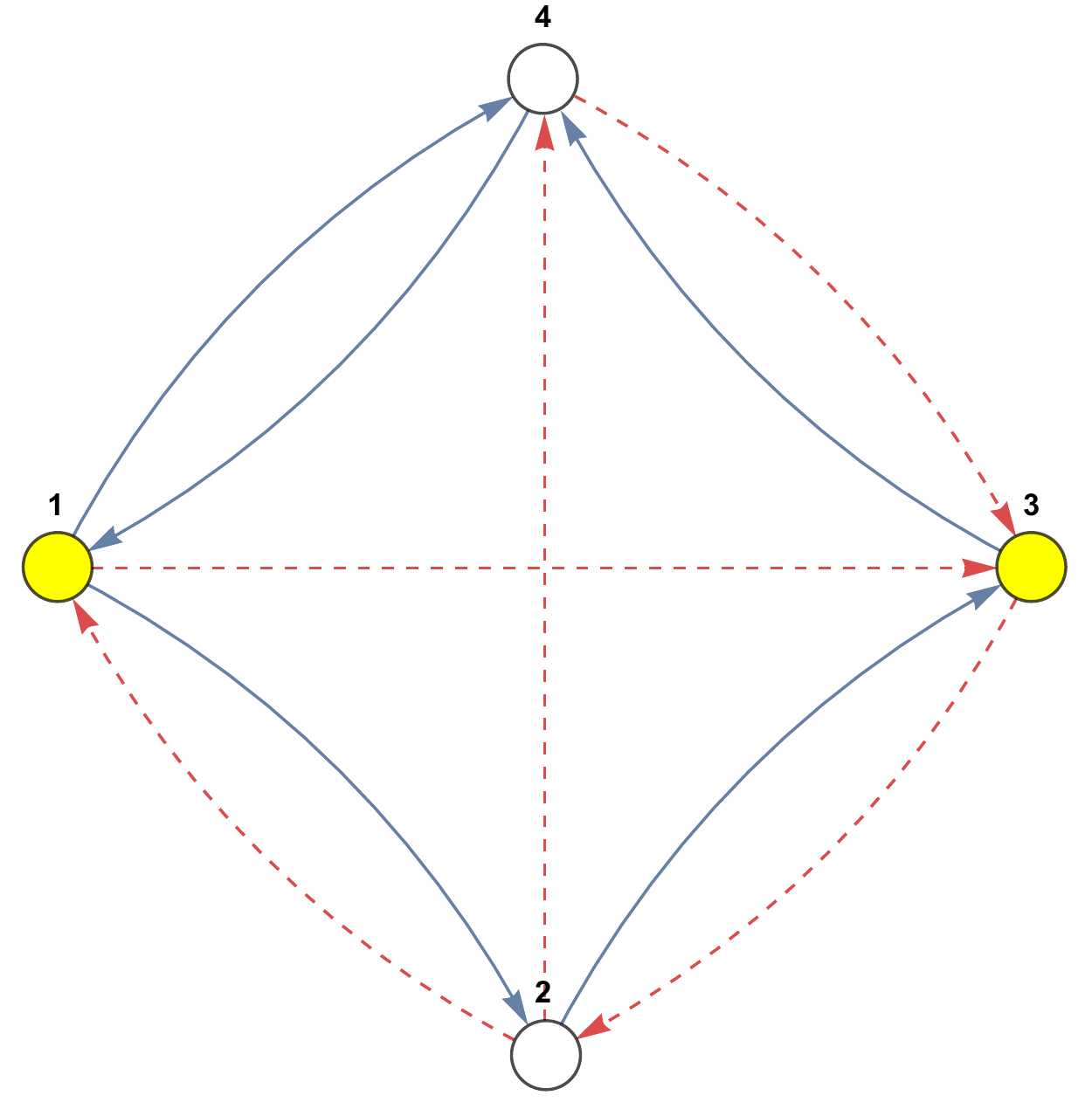} 
\end{minipage}
& 
\begin{minipage}{0.33\textwidth}
\centering
\begin{tabular}{|>{\columncolor[gray]{0.8}} c | >{\columncolor[gray]{0.8}} c |}
\hline
0 1 0 1 & 0 0 1 0 \\
0 0 1 0 & 1 0 0 1 \\
0 0 0 1 & 0 1 0 0 \\
1 0 0 0 & 0 0 1 0 \\
\hline
\end{tabular}
\end{minipage}
& 
\begin{minipage}{0.33\textwidth}
\centering
$
\xymatrixrowsep{0.1cm}
\xymatrixcolsep{0.5cm}
\xymatrix{
1 0 1 0 & 1 0 1 0 \\
0 1 0 1 & 0 1 1 0 \\
1 0 1 0 \ar@(l ,l)@[blue][uu] & 1 1 0 1 \\
\hdots & 1 0 1 1 \\
& 0 1 1 0 \ar@(r ,r)@[red][uuu]\\
& \hdots 
}$
\end{minipage}
\end{tabular}%
\end{table}

Which attracting set a state is part of (or whether it is a transient state) is determined by the network topology and the rules governing its evolution. In our toy model, the same initial state $(1,0,1,0)$ reaches two different attracting sets (indicated by the colored arrows in Table \ref{tab:one}). In the blue case, the initial state is part of an attracting set of size 2, whereas in the red case it is transient. In the blue case, the network realizes a flip-flop circuit which permanently oscillates between the states $(1,0,1,0)$ and $(0,1,0,1)$ (Table. \ref{tab:one}, right column, left-hand -side). In the red case, the network falls into an attracting set which cycles through its states with period 3 (Table. \ref{tab:one}, right column, right-hand -side). In both cases, an initial state of $(0,0,0,0)$ would be stationary and could be interpreted as ground- or terminal state of the network.

The stationary states and attracting sets represent (``kinetically'') stable entities which are associated with the network. This gives an explicit answer to the question how a (transitory) ontology on the individual level could give rise to stable entities on the level of networks. We will now show how such states could also be conceived of as encoding perceptual inferences. 

\subsection{Perceptual Inference}
Mathematically, the process of perception is often modeled as abductive inference using Bayesian probability theory. The perceptual system chooses from a variety of ``interpretations'' $x$, each consistent with an input signal $i$. A posterior probability can be assigned to each interpretation, given the input. Using Bayes' rule, it can be recovered from multiplying a likelihood function that assigns a probability to each input given a particular percept with the prior probability of that percept:
\begin{equation}
p(x|i) =p(i|x) \cdot \frac{p(x)}{p(i)} =\alpha \cdot p(i|x)p(x), 
\end{equation}
where $\alpha=p(i)=\sum_x p(i|x)p(x)$  is a normalization constant, which can be neglected when comparing different posterior probabilities given the same input. The way how this scheme is (approximately) realized in a living organism can be quite complicated. 

In vision science, it is often assumed that the space of percepts is homomorphic to the space of states in the world, and thus the likelihood function could be given the interpretation of ``a mapping'' from world states to input states, e.g., the optical projection from a 3D world onto the retina of the eye. Such an approach to visual perception is sometimes called an ``inverse-optics''  approach, since the task for the visual system is to undo the effects of optical projection \cite{Pal99}. 
Given such an interpretation and knowing likelihood and prior probability, it is possible to calculate posterior probabilities using the Bayesian scheme above. To determine which percept $x$ is selected based on the posterior distribution, one usually introduces a loss function that describes the error in the process of choosing an interpretation. The most straightforward and principled loss function is given by a delta-function, centered around the maximum of the posterior distribution (the so-called ``MAP'' - estimate). Other, more involved choices are possible \cite{Mam02}. 
In an evolutionary setting, the assumption that world states and scene interpretations generically are homomorphic has shown to be too strong an assumption \cite{Hof15,Mar10,Pra20}, and we thus refrain from it. 

For a \textsf{perceptual network} we assume that the ``sensory input'' $i$ corresponds to the initial state of a certain (sub-)set of nodes and ``perceptual interpretations'' $x$ correspond to the stable states of the network. 
The set of rules $\{ R_j \}$ encodes a ``perceptual strategy'', and knowing the rules and the network’s topology, we could compute posterior probabilities directly based on priors defined on inputs to maximize a quantity that represents a ``fitness payoff'' with respect to the problem at hand.  We thus introduce a ``payoff-function'' defined on the space of inputs and perceptions:
\begin{equation}
F :I\times X \rightarrow [0,1],\ \ (i,x)\mapsto r\in [0,1]. 
\end{equation}
What the network sees is not merely ``given'' by its input but by the the way it is internally processed, reflected by the utility of its strategy. On this account, the evolution of perceptions does not serve as ladder to the truth but as way to find strategies that maximize fitness payoffs. 

For convenience, one often chooses the interval $[0,1]$ for payoff values, but in general any finite interval defined on the positive Reals will suffice. 
With this definition of a payoff-function we could calculate an ``expected payoff":
\begin{equation}
\langle F \rangle = \sum_{i,x} F(i,x) p_A(i,x| {R_j }) 
\end{equation}
We want to infer the ``average best'' perceptual system implemented by a network. (For simplicity, we chose to keep the topology of the graph fixed which is indicated by $p_A$.) Since the posterior probability $p_A(x|i)$ is determined by the set of rules $\{R_j\}$, we need to adjust the rules in order to maximize the expected payoff which is related to the state of our network:

\begin{equation}
\label{eq:eight}
\{R_j\}_\textrm{opt} = \textrm{argmax}\left( \langle F \rangle \right) = \textrm{argmax} \left( \sum_{i,x,} F(i,x)p_A(i,x|\{R_j\}) \right)
\end{equation}
From now on we will drop the conditional dependence on $\{R_j\}$ for simplicity when writing fitness payoffs.

\subsection{Mutual Information}
In our model, and in absence of any further assumption, we assume a particular form of the payoff function which equals the logarithm of the probability $p_A(i,x)$ divided by the probability of the marginal probabilities $p(i)p_A(x)$. This results in a mutual information:
\begin{eqnarray}
\langle F_A \rangle = \MI{I}{X}_A &=&\sum_{i,x} p_A(i,x) \log \left( \frac{p_A(i,x)}{p(i)p_A(x)} \right) \\
& = & \sum_i p(i)\cdot \DKL{p_A(x|i)}{p_A(x)} =  \langle \DKL{p_A(x|i)}{p_A(x)} \rangle_i .
\end{eqnarray}
In other words, the payoff-function is set equal to the difference in self-information between a posterior probability $p_A(x|i)$ and a marginal probability $p_A(x)$. More intuitively, $\langle F_A \rangle$  could be thought of as quantifying the information that is present in the asymptotic states of the \textsf{perceptual network} about the initial state.  
The maximum value of the mutual information is bounded by the entropy $H(I)$ of the input:
\begin{equation}
\textrm{max}(\MI{I}{X}) \le \textrm{max}(H(I)).
\end{equation}
This bound is obtained whenever the perceptual rule leads to a one-to-one mapping from inputs to perceptions. In any realistic setting (e.g. where the space of inputs largely exceeds the state of percepts) this will likely not be the case, and we have to relate input and perception probabilistically and optimize Eq. \eqref{eq:eight} instead. 

Already a noisy input would force us to regard this as a stochastic problem which will generally lower the maximal value of mutual information achievable in the network. To each sensation there exist several perceptual representations which all have a certain posterior probabilities, which are specified by the network's evolution under noise (simply speaking, ``noise'' is taken to lead to a perceptual ``misrepresentation'' of the input). 

On this account the ``goal'' of perception is being able to distinguish as best as possible between different inputs. Perception, thus understood, amounts to the ability to recognize differences in the world (or more generally: perception is the ability to recognize ``differences that make a difference'' \cite{Bat79}  for fitness). Finding a set of rules defined for individual nodes which maximizes the mutual information between perception and input of the network is the ``perceptual problem'' that needs to be solved by our model. For this, we can borrow techniques to search through the rule-set using algorithms from machine learning. One possibility is to use evolutionary programming techniques such as genetic algorithms \cite{Hol92} as discussed below.%
\footnote{The formulation of the problem makes it in principle amenable to optimization techniques involving calculating gradients of a ``free-energy'' functional (e.g. \cite{Fri10}). However, in the more general setup where arbitrary (non-differentiable and non-continuous) payoff functions are used, such methods are no longer straightforwardly applicable.}

\section{A preliminary experiment}

\subsection{Models and Methods}
\label{sec:methods}
We tested the performance of a \textsf{perceptual network} with $N=16$ nodes for 3-bit inputs distributed on the first 3 nodes. The system was initialized in the state $X=(i[1],i[2],i[3],0,…,0)$ and was evolved according to a set of rules $\{ R_j \}$ on a fixed topology $A$. The attracting sets were recorded for each input. This allowed us to construct a conditional probability $p(x|i)$ defined on the space of inputs, which later informed mutual information (i.e. fitness payoffs). 

For each input, we introduced uncertainty by including a probability ($p=0.05$) to flip an input bit during the initialization step. We then used a genetic algorithm (GA) to determine the best set of rules in terms of fitness. Each rule is specified by the integer values $\mu_1$ and $\mu_2$ (see Eq. \eqref{eq:ten} below), which thus defined the ``genes'' of a \textsf{perceptual network}. We evolved the network for $50$ generations, each comprising 25 individuals. The best 20\% were kept ($\lambda =0.2$), the remaining 80\% were generated using a fitness-weighted sexual recombination procedure. We included a fixed mutation probability ($p=0.01$) of randomly shifting the values of $\mu_1$ and $\mu_2$ by  $\pm 1$. To make our model biologically more plausible, we assumed a refractory period of 1, which means that any node – after it has been excited – cannot be excited immediately in the next round. 

We chose a rule that assigns the next state of a node solely dependent on the number of messages it receives, independent of source or any other statistical property. More concretely, we assume a rule of the following type:
\begin{equation}
\label{eq:ten}
x_j '=
\begin{cases} 
1 \textrm{ if } m_j \in [\mu_1,\mu_2] \textrm{ and } x_j =0\\
0 \textrm{ else }
\end{cases},
\end{equation}
with $1 \le \mu_1 \le \textrm{deg}(j)$ and $\mu_1 \le \mu_2 \le \textrm{deg}(j)$  where the messages $m_j$ are computed as $m_j = (\textbf{A}\vec{x})_j$ (cf. the simple example in section \ref{sec:example} ). Any node thus receives a message which is the sum of all states of the adjacent nodes.
In total there are $\frac{1}{2}\cdot \textrm{deg}(j) \left[\textrm{deg}(j)+1 \right]$ possible rules per node with degree $j$. If we consider a network of size $N$, each node independently following a different rule, the total number of rules are:
\begin{equation}
N_R = \prod_{j=1}^N \frac{\textrm{deg}(j) ( \textrm{deg}(j)+1)}{2}  
\approx  \left( \frac{ \langle \textrm{deg}\rangle^{2} } {2} \right)^N.
\end{equation}
(Assuming that each node has approximately the same average degree $\langle \textrm{deg} \rangle$.) So, even though the rule expressed in Eq. \eqref{eq:ten} is quite simple, the rule space scales exponentially with the size of the network. 

In this investigation, we optimized the set of rules $\{ R_j \}$ for a fixed network topology which either resembled (i) a 4-lattice, (ii) a scale free (SF) network  constructed using the Barab\'asi-Albert method \cite{Bar02} , or (iii) a complete graph where any node is connected to any other other node. We also tested our results against randomly generated (Erd\"os-Renyi) networks with sparse ($\langle \textrm{deg} \rangle \approx \frac{N}{10}$) and dense ($\langle \textrm{deg} \rangle \ge \frac{N}{2}$) degree distributions.

We compared the evolved rule-set to a variant of a majority rule where each node transitions from 0 to 1 if and only iff at least half of its neighbors are in state 1, else it goes to 0. The majority rule has previously been found to be very efficient in solving the ``density classification task'' for a cellular automaton when defined on small-world graphs \cite{Str98}. The density classification task is highly global but involves only a single rule at the local level. It could thus be regarded as benchmark test for local and parallelized computational architectures. 

Note that one could rely on existing Python libraries for using genetic algorithms (e.g. \cite{deap}) or use other optimization methods than evolutionary ones, e.g. graph neural networks \cite{Sci09}. We decided to implement a simple genetic algorithm by hand instead; our goal was sketching the conceptual ideas of \textsf{perceptual networks}.

%Another fixed rule which we used for comparison is the ``max-thresh'' rule, where a node is updating iff all its neighbors  
%So, for example, if an agent has 3 other neighbors to communicate with, the state transition takes place if and only if |m_j |=3 etc. 

\subsection{Results and Discussion}

Results for some exemplary networks are given in Table \ref{tab:two}. The rules which evolved after $50$ generations of the GA, favor interface strategies \cite{Hof15} , which generically lead to no structural similarity between input states and asymptotic states of the network. The perceptual states of the network do not mirror any structure in the input other than a probabilistic relationship given by the posterior $p(x|i)$ which informed fitness payoff (mutual information). 

By contrast, the majority rule mirrors the structure of the messages.  More precisely, the majority rule says that the state of a node is a homomorphic representation of the messages it receives: the more content in the message, the higher the probability to undergo state transition (in our implementation either yes/no). While the majority rule led to satisfying solutions for particular topologies, it is generically not able to compete with a collection of interface rules $\{R_j\}$ on a random topology.

Different behaviors were observed for different topologies: SF networks lead to a quick emergence of ``fit'' strategies on a network with only sparse degree distribution ($\langle \textrm{deg} \rangle < 2$). It is expected that such networks have comparatively low computational cost in a realistic case where connections are expensive. 
For the 4-lattice we found that the average fitness payoff increased more slowly compared to SF networks, although it generally gave good results, depending on the initial configuration of the network. The average degree distribution is much higher ($\langle \textrm{deg} \rangle = \sqrt{N}$). 
Good and fast converging results have been obtained for complete graphs due to the intrinsic refractionory period of the rule. Since only a few combinations of rules lead to good results, convergence time has been small. However, we expect the estimated computational cost in realistic scenarios to be much higher for such topologies (due to the average degree of $\langle \textrm{deg} \rangle \approx N$). SF networks thus promise to offer a good compromise between cost and fitness. 

Similar results have been obtained for randomized graphs, where networks with sparse degree distribution behaved similar to the SF networks and networks with dense degree distributions behaved similar to complete networks. This indicates that the decisive factor is average degree distribution. But effects for different placements of the initial nodes have been observed in the SF case (roughly: networks where input nodes lie in different communities of the graph, behave optimally)

The results have not been assessed rigorously in terms of statistics and should only convey a preliminary sense of what is possible with this method of inquiry.  

\begin{table}
\centering
\caption{Results for some randomly initialized networks, defined on a fixed topology visualized in the left column (input states are highlighted in red). The perceptual rules for the fittest strategy after the genetic algorithm are displayed in the middle column by the thresholds $\mu_1$ and $\mu_2$ for each node. Averaged values for mutual information (fitness) for the initial and final population after the GA and results for the majority are given on right column (the maximally achievable value is 2.14.) }
\begin{tabular}{lll}
\toprule
Network, $N=16$ & Thresholds & Mutual information (std. dev)\\\hline
\begin{minipage}{0.43\textwidth}
4-lattice, $\langle \textrm{deg} \rangle = 4$\\
\includegraphics[scale=0.75]{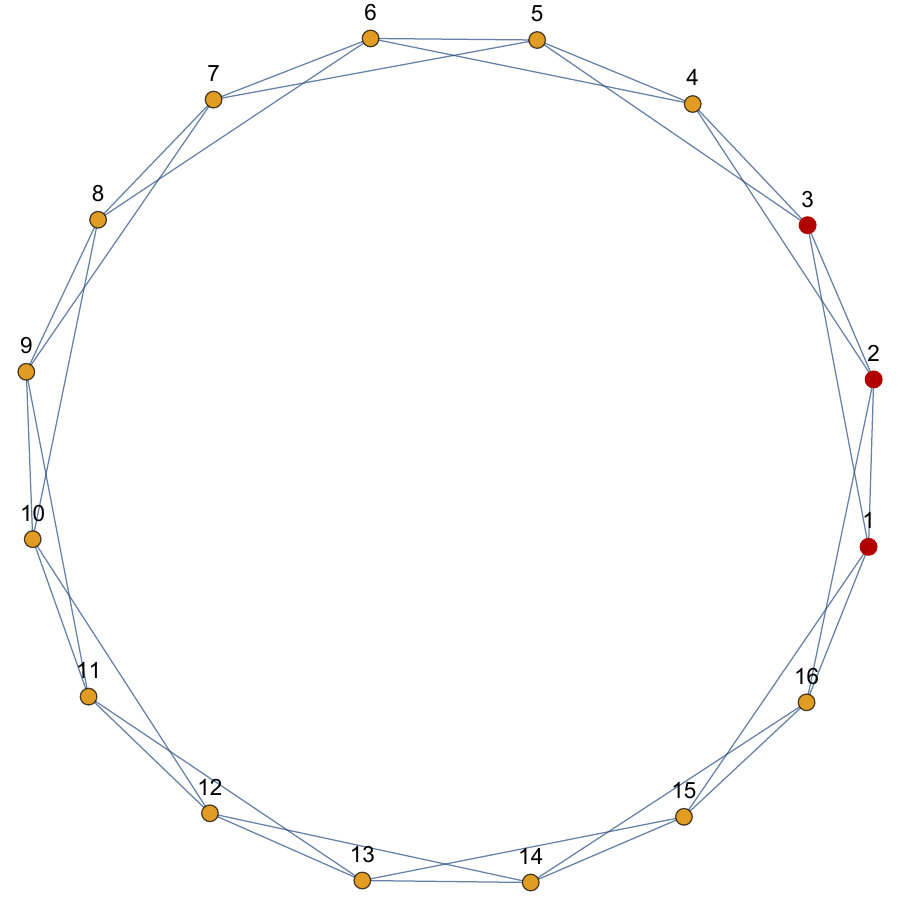}
\end{minipage}
& 
\begin{minipage}{0.26\textwidth}
\centering
$
\mu_1= \begin{pmatrix}
1 \\ 4 \\ 2 \\ 3 \\ 1 \\ 5 \\ 1 \\ 1 \\ 1 \\ 2 \\ 1 \\ 1 \\ 2 \\ 1 \\ 1 \\ 1 
\end{pmatrix}
,
\mu_2 = \begin{pmatrix}
2 \\ 5 \\ 2 \\ 3 \\ 5 \\ 6 \\ 5 \\ 5 \\ 4 \\ 2 \\ 4 \\ 2 \\ 4 \\ 2 \\ 3 \\ 5
\end{pmatrix} 
$
\end{minipage}
&
\begin{minipage}{0.31\textwidth}
\centering
\begin{tabular}{c|c|c}
$\MI{I}{X}_0$ & $\MI{I}{X}_{50}$ & $\MI{I}{X}_{\textrm{maj}}$ \\\hline
0.22 (0.34) & 1.92 (0.33) & 0.70 (0.0)
\end{tabular}
\end{minipage}
\\
\begin{minipage}{0.43\textwidth}
Scale-free network (m=1), $\langle \textrm{deg} \rangle = 1.88$\\
\includegraphics[scale=0.75]{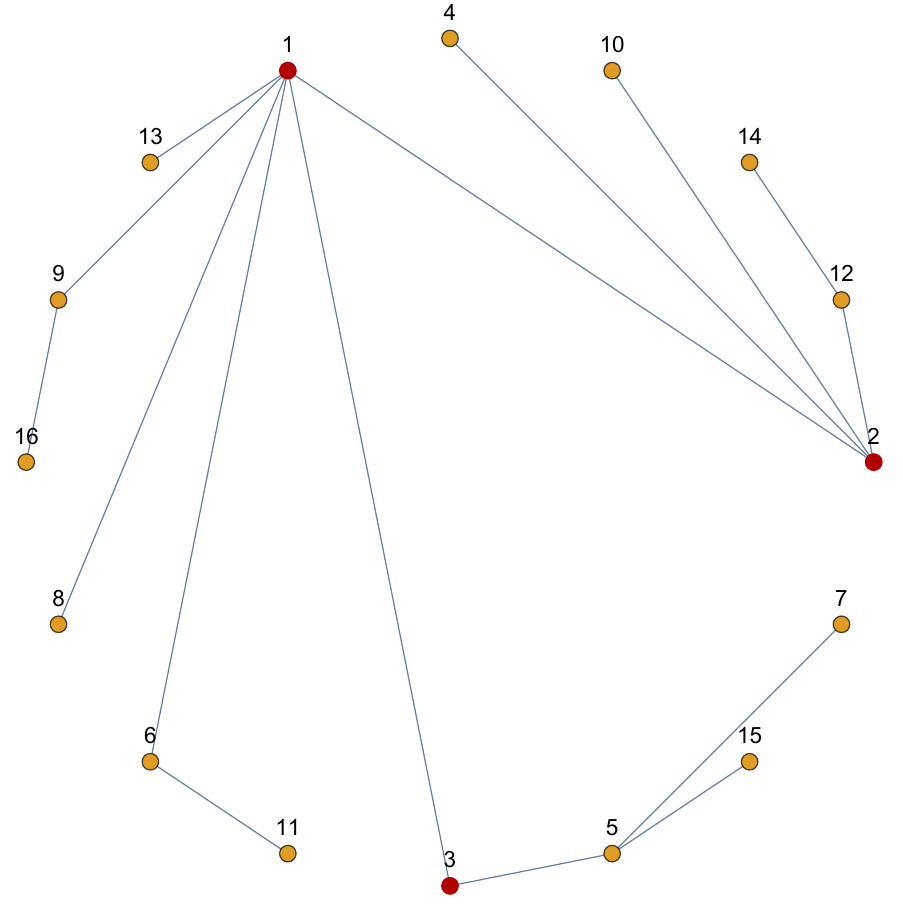}
\end{minipage}
& 
\begin{minipage}{0.26\textwidth}
\centering
$
\mu_1= \begin{pmatrix}
4 \\1 \\ 3 \\ 1 \\ 1 \\ 1 \\ 2 \\ 2 \\ 1 \\ 2 \\ 1 \\ 3 \\ 2 \\ 2 \\ 1 \\ 2 
\end{pmatrix}
,
\mu_2 = \begin{pmatrix}
5 \\ 3 \\ 4 \\ 2 \\ 2 \\ 3 \\ 2 \\ 2 \\ 3 \\ 2 \\ 2 \\ 3 \\ 2 \\ 2 \\ 1 \\ 2
\end{pmatrix} 
$
\end{minipage}
&
\begin{minipage}{0.31\textwidth}
\centering
\begin{tabular}{c|c|c}
$\MI{I}{X}_0$ & $\MI{I}{X}_{50}$ & $\MI{I}{X}_{\textrm{maj}}$ \\\hline
0.7 (0.37) & 1.74 (0.35) & 1.43 (0.0)
\end{tabular}
\end{minipage}
\end{tabular}
\label{tab:two} 
\end{table} 
\begin{table}
\captionsetup{singlelinecheck = false, justification=justified}
\caption*{Table 2 - continued}
%{Results for some randomly initialized networks (continued), defined on a fixed topology visualized in the left column (input states are highlighted in red). The perceptual rules for the fittest strategy after the genetic algorithm are displayed in the middle column by the values of $\mu_1$ and $\mu_2$ for each agent. Averaged values for mutual information (fitness) for the initial and final population after the GA and results for the majority are given on right column (the maximally achievable value is 2.14.) }
\begin{tabular}{lll}
\toprule
Network, $N=16$ & Thresholds & Mutual information (std. dev)\\\hline
%\\
\begin{minipage}{0.43\textwidth}
Complete graph, $\langle \textrm{deg} \rangle = 15$\\
\includegraphics[scale=0.75]{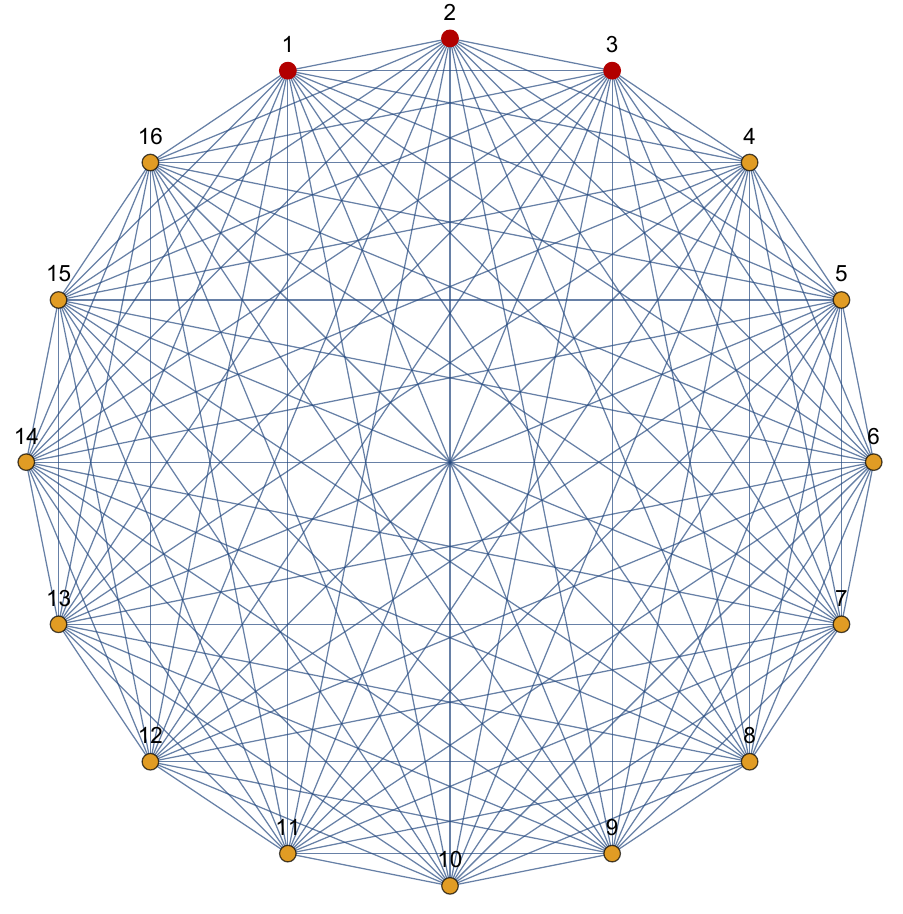}
\end{minipage}
& 
\begin{minipage}{0.28\textwidth}
\centering
$
\mu_1= \begin{pmatrix}
1 \\ 2 \\ 2 \\ 4 \\ 13 \\ 4 \\ 7 \\ 8 \\ 2 \\ 15 \\ 14 \\ 1 \\ 6 \\ 8 \\ 3 \\ 9
\end{pmatrix}
,
\mu_2 = \begin{pmatrix}
3 \\ 3 \\ 7 \\ 16 \\ 14 \\ 14 \\ 13 \\ 15 \\ 8 \\ 15 \\ 14 \\ 16 \\ 13 \\ 16 \\ 14 \\ 16
\end{pmatrix} 
$
\end{minipage}
&
\begin{minipage}{0.329\textwidth}
\centering
\begin{tabular}{c|c|c}
$\MI{I}{X}_0$ & $\MI{I}{X}_{50}$ & $\MI{I}{X}_{\textrm{maj}}$ \\\hline
0.33 (0.15) & 1.83 (0.20) & 0.0 (0.0)
\end{tabular}
\end{minipage}
\end{tabular}
\label{tab:two-cont} 
\end{table} 

\section{Conclusion and Outlook}
In the model outlined in the previous section, we have modeled perceptual inference in terms of strategies that compute the mutual information between input and asymptotic states of a network. In general, we found that the use of genetic algorithms leads to quick convergence on interface strategies, i.e. strategies that do not follow any fixed (non-evolvable) rule which, generically, do not mirror the structure of the input. 

While we have optimized rules of the individual nodes, the topology of the network has been fixed. In a next experiment, topology should be optimized as well.  First, realize that each column of the adjacency matrix could be interpreted as an ``action'' that specifies message constructing for each node at each updating step. Analogously to how the evolution of rule sets could be interpreted as an evolution of how the network ``perceives'' information, the evolution of topology could be interpreted to an evolution of this ``action'' (on itself). Similar as before, a genetic algorithm could be used to update the adjacency matrix, where each column is encoded by a ``chromosome'' associated to each node of the network with ``genes'' representing the locations where future messages will be sent to. Occasionally during the recombination process, a random mutation could flip a bit (i.e. a gene), that is, create or destroy a new link in the adjacency. This means that the target of the evolutionary process would be the genetic information carried at each node.

Alternatively, one could implement the method reported in \cite{Str98} that aims at producing a small world network based on a locally dense initial graph (such as the 4-lattice) through occasionally rewiring one link (and thus turning it from a local connection to a global connection). 

%Second, instead of undirected networks, as used above, one could consider directed graphs, where the direction of each arrow specifies the direction of communication. 

What would one expect to happen? As previously indicated, scale free topologies present a good compromise between fitness and computational cost and are thus expected to be the favored outcomes of a learning (evolutionary) process which optimized the topology of the network under the constraint of keeping the average degree small. 
Such networks hint at the presence of important sub-networks of different sizes, i.e. ``hubs'' which integrate a lot of incoming information and/or send a lot of messages into the network. It is furthermore likely that a single node on a coarser level is itself composed of many interacting sub-networks (``hubs within hubs''). This would indicate that the model is representing a hierarchy of (heterogenous) agents, a fact which is, however, observer-relative and depends on the coarse-graining chosen (on the elementary level the network is just a collection of 1-bit nodes).

A further developed model could be conceived along several lines: (i) Use a more sophisticated – and heterogenous – model of the refractory dynamics with varying periods between excitable states. (ii) Use a non-synchronous updating scheme. Ideally, the network would feature a combination between asynchronous updating and synchronous updating schemes for several groups of nodes. (iii) Introduce probabilistic updating. In our model, probabilistic inference was restricted to initial uncertainty, the strategies and topologies where otherwise deterministic. The natural interpretation of probabilities in the adjacency regards the interaction probability between adjacent nodes: Edges with probability close to 1 represent nodes that almost always exchange signals, whereas edges with probability close to 0 represent nodes that almost never communicate. In a probabilistic setting with low fluctuations around a central value, the network topology might play an even bigger role than in the previous subsection. We conjecture that SF networks are robust with respect to small fluctuations, as has been demonstrated for many empirical networks which approximate such a topology \cite{Bar02}.  This would distinguish SF networks further others, e.g., lattice structures.

\clearpage

\end{document}